\documentstyle[12pt]{article}

\begin{document}

\centerline{{\Large\bf The Moyal bracket and the dispersionless}}

\vspace{.2in}
\centerline{{\Large\bf limit of the KP hierarchy}}

\vspace{.2in}
\centerline{{\bf I.A.B.Strachan}}

\vspace{.1in}
\centerline{Dept. of Mathematics and Statistics, University of Newcastle,}
\vspace{.1in}
\centerline{Newcastle-upon-Tyne, NE1 7RU, England
\footnote{e-mail: i.a.b.strachan@newcastle.ac.uk}.}

\vspace{.3in}
\centerline{{\bf Abstract}}

\vspace{.2in}
\small
\parbox{5.8in}{A new Lax equation is introduced for the KP hierarchy
which avoids the use
of pseudo-differential operators, as used in the Sato approach. This Lax
equation is closer to that used in the study of the dispersionless KP
hierarchy, and is obtained by replacing the Poisson bracket with the Moyal
bracket. The dispersionless limit, underwhich the Moyal bracket collapses
to the Poisson bracket, is particularly simple.}
\normalsize

\bigskip
\section*{1. Introduction }

One of the simplest nonlinear equations that can be completely be solved,
albeit implicitly, is

\[
4 U_T - 12 U U_X = 0 \,,
\]

\noindent the solution to which can be obtained using the method of
characteristics. This equation
can be described in two ways, either as the dispersionless KdV equation (i.e.
the KdV equation without the dispersion $U_{XXX}$ term) or as the simplest
example of an equation of hydrodynamic type. This connection between
dispersionless and hydrodynamic equations persists into the
theory of $(2+1)$-dimensional systems. The dispersionless KP equation
(hereafter referred to as the dKP equation)

\[
(4 U_t - 12 U U_X )_X  = 3 U_{YY}
\]

\noindent may be obtained from the KP equation itself

\[
(4 u_t - 12 u u_x - u_{xxx} )_x = 3 u_{yy}
\]

\noindent via the scaling transformation

\begin{equation}
\begin{array}{rcl}
X & = & \epsilon x \,, \\
Y & = & \epsilon y \,, \\
T & = & \epsilon t \,, \\
U(X,Y,T) & = & u(x,y,t)
\end{array}
\end{equation}

\noindent in the limit as $\epsilon \rightarrow 0\,.$ The dKP may be reduced to
the
hydrodynamic-type equation

\[
{\partial {\bf w} \over \partial T} = {\bf A}({\bf w})
{\partial {\bf w} \over \partial X}\,,
\]

\noindent where ${\bf A}$ is an $N\times N$ matrix and ${\bf w}$ is an
$N$-component column vector ($N,$ which characterises this reduction,
is an arbitrary integer). Such a  reduction enables one to construct solutions,
and study the properties of, the dKP equation by using existing results on
the theory of equation of hydrodynamic type {\bf [1]}.

\bigskip

The dKP and KP equations are important examples of $(2+1)$-dimensional
integrable systems, both having associated Lax equation. For the KP
equation (or more generally, for the KP hierarchy) the Lax equation
is {\bf [2]}

\begin{equation}
{\partial {L} \over \partial t_n} =
\Big[  ({L}^n)_{+}, {L} \Big]\,,
\end{equation}

\noindent where $\partial = {\partial\phantom{x}\over\partial x}\,,$

\[
{L} = \partial + \sum_{n=2}^{\infty} v_n(x,t_2,t_3,\ldots)
\partial^{-n+1}
\]

\noindent and $\Lambda_{+}$ denotes the projection onto the differential
operator part of the pseudo-differential operator $\Lambda\,.$ The bracket
$[A,B]$ is just the commutator of the differential operators, i.e.
$[A,B] = A\,B - B\,A\,.$

\bigskip

The Lax equation for the dKP hierarchy is somewhat different, as it does not
involve the use of pseudo-differential operators. The Lax equation is {\bf [1]}

\begin{equation}
{\partial {\cal L} \over \partial t_n} =
\Big\{  ({\cal L}^n)_{+}, {\cal L} \Big\}\,,
\end{equation}

\noindent where

\[
{\cal L} = \lambda + \sum_{n=2}^{\infty} u_n(x,t_2,t_3,\ldots)
\lambda^{-n+1}
\]

\noindent and $\Omega_{+}$ denotes the projection onto positive (and zero)
powers of $\lambda$ in the Laurent series $\Omega\,.$ The bracket $\{f,g\}$
is just the Poisson bracket

\begin{equation}
\{f,g\} =
{\partial f\over \partial \lambda} {\partial g \over \partial x} -
{\partial g\over \partial \lambda} {\partial f \over \partial x}\,.
\end{equation}

\bigskip

One interesting point to notice is that, although the scaling transformation
takes one from the KP equation to the dKP equation, if one applies it to the
Lax equation for the KP equation one does not obtain the Lax equation for the
dKP equation, at least not in any naive way. This may be summarised as the
failure of the following diagram to commute:

\halign{\hskip 15mm\indent\quad\quad\hfil#\hfil&\hfil#\hfil&\hfil#\hfil\cr
 &\phantom{~~~~~~~~~~~~~~~~~~}& \cr
KP hierarchy&$\longleftrightarrow$  & Lax equation (2) \cr
\noalign{\vskip 1mm}
{\phantom{scaling (1) }}$\vert$~~~~~~~~~~~~&~~&$\vert$\cr
\noalign{\vskip -1mm}
{\phantom{scaling (1) }}$\vert$~~~~~~~~~~~~&~~& \cr
\noalign{\vskip -1mm}
scaling (1) $\vert$~~~~~~~~~~~~&~~&$\vert$\cr
\noalign{\vskip -1mm}
{\phantom{scaling (1) }}$\vert$~~~~~~~~~~~~&~~&\cr
\noalign{\vskip -1mm}
{\phantom{scaling (1) }}$\downarrow$~~~~~~~~~~~~&~~&$\downarrow$\cr
\noalign{\vskip 1mm}
dKP hierarchy&$\longleftrightarrow$&Lax equation (3)\cr}

\bigskip

\bigskip

\noindent The aim of this letter is to introduce a new Lax equation for the KP
hierarchy so that the above diagram does commute. In fact one will obtain
the KP equation in the form

\[
(4 u_t - 12 u u_x - 4 \kappa^2 \, u_{xxx} )_x = 3 u_{yy}
\]

\noindent and so the dispersionless limit corresponds to
$\kappa\rightarrow 0\,.$ This then avoids the scaling transformation. This
will be achieved by replacing the Poisson bracket in (3) by the Moyal
bracket, and the dispersionless limit is the limit in which the Moyal
bracket collapses to the Poisson bracket.

\section*{2. The Moyal Bracket}

The Moyal bracket {\bf [3]} may be thought of as a deformation of the Poisson
bracket
by the introduction of higher order derivative terms. It turns out that the
Jacobi identity is highly restrictive as to the nature of these terms, and one
is lead uniquely {\bf [4]} to the Moyal bracket:

\begin{equation}
\{f,g\}_\kappa = \sum_{s=0}^{\infty} { \kappa^{2s} \over (2s+1)! }
\sum_{j=0}^{2s+1} (-1)^j
\left( \begin{array}{c} 2s+1 \\ j \end{array} \right)
(\partial_x^j \partial_\lambda^{2s+1-j} f )
(\partial_x^{2s+1-j} \partial_\lambda^j g )\,.
\end{equation}

\noindent It has all the standard properties one would expect of such a
bracket:

\[
\begin{array}{lll}
{\rm (a)~~}&\{f,g\}_\kappa = - \{f,g\}_\kappa\,,&{\rm antisymmetry\,,} \\
{\rm (b)~~}&\{af+bg,h\}_\kappa = a\{f,h\}_\kappa + b\{g,h\}_\kappa\,, ~~~ &
{\rm linearity\,,} \\
{\rm (c)~~}&\{f,\{g,h\}_\kappa\}_\kappa + \{g,\{h,f\}_\kappa\}_\kappa +
\{h,\{f,g\}_\kappa\}_\kappa = 0\,, &{\rm Jacobi~identity}
\end{array}
\]

\noindent (where $a,b$ are independent of $x$ and $\lambda\,$). Moreover it
has the important property that

\begin{equation}
\lim_{\kappa\rightarrow 0} \, \{f,g\}_\kappa =\{f,g\}\,,
\end{equation}

\noindent i.e. in the limit as $\kappa\rightarrow 0$
the bracket collapses to the Poisson bracket (4). It also has many other
interesting properties {\bf [5]}, amongst which is the fact that it may be
written in
terms of an associative $\star$-product defined by

\[
f\star g = \sum_{s=0}^{\infty} {  \kappa^{s} \over s! }
\sum_{j=0}^{s} (-1)^j
\left( \begin{array}{c} s \\ j \end{array} \right)
(\partial_x^j \partial_\lambda^{s-j} f )
(\partial_x^{s-j} \partial_\lambda^j g )\,.
\]

\noindent  This has the property that

\begin{equation}
\lim_{\kappa\rightarrow 0 } f \star g = fg\,,
\end{equation}

\noindent and with this the Moyal bracket takes the form:

\[
\{f,g\}_\kappa = {f\star g - g\star f\over 2\kappa}\,.
\]

\bigskip

The hierarchy to be considered here is obtained by replacing the Poisson
bracket in (3) by the Moyal bracket:

\begin{equation}
{\partial {\cal L}\over \partial t_n} = \{ {\cal B}_n , {\cal L} \}_\kappa\,,
\end{equation}

\noindent where

\[
{\cal B}_n =
( \underbrace{ {\cal L} \star \ldots \star {\cal L} }_{\rm n-terms} )_{+}
\]

\noindent and ${\cal L}$ remains unchanged. This is equivalent to
the zero-curvature relations

\begin{equation}
{\partial {\cal B}_n \over \partial t_m}-
{\partial {\cal B}_m \over \partial t_n}+
\{ {\cal B}_n,{\cal B}_m\}_\kappa = 0 \,.
\end{equation}

\bigskip

\noindent Since the ${\cal B}_i$ are
all polynomial in $\lambda$ the Moyal bracket will automatically truncate
after a finite number of terms, and so one obtains a well defined
set of evolution equation for the independent variables. These equations
differ from those of the dKP hierarchy by a finite number of $\kappa$-dependent
terms. From (8) and (9) one may prove a number of general properties of the
hierarchy. For example, using equations (6) and (7), as $\kappa\rightarrow 0$
the hierarchy reduces to the dKP hierarchy.

\bigskip

\noindent {\bf Example}

\bigskip

With  $n=2$ and $n=3$ equation (8) yields

\[
\begin{array}{rcl}
{\cal B}_2 & = &  \lambda^2  + 2 u_2 \,, \\
{\cal B}_3 & = & {\lambda^3} + 3 \lambda u_2 + 3 u_3 \,
\end{array}
\]

\noindent ($\kappa$-dependent terms only appear in ${\cal B}_n$ for $n>3$),
and one obtains from (9):

\[
\begin{array}{rcl}
-3  u_{2,t_2} + 6 u_{3,x} & = & 0 \,, \\
2u_{2,t_3} - 6 u_2 u_{2,x} -2 \kappa^2 u_{2,xxx} - 3 u_{3,t_2}& = & 0 \,.
\end{array}
\]

\noindent On eliminating $u_3$ one obtains a single equation for $u_2\,:$

\[
(4 u_{2,t_3} - 12 u_2 u_{2,x} - 4 \kappa^2 u_{2,xxx})_x = 3 u_{2,t_2 t_2}\,,
\]

\noindent the KP equation itself. For this to agree with the KP equation
obtained from (2) one has to set $\kappa^2 = \frac{1}{4}\,.$ Further, as
$\kappa\rightarrow 0$ one obtains the dKP directly without the need of the
scaling transformation (1).

\bigskip

\bigskip

\bigskip

It therefore seems plausible that (8) is the KP hierarchy, and this has
been proved for the first four members of the hierarchy by direct calculation.
However, a direct proof that (8) is equivalent to the hierarchy given by the
Sato approach is lacking at present, though it does seem highly
unlikely that in addition to having the same dispersionless limit as the
KP hierarchy, it should agree with the KP hierarchy for the first four terms
and then, after that, diverge. An additional problem is that the functions
$v_n$ and $u_n$ appearing in $L$ and ${\cal L}$ are not identical, but are
related by simple relations, the first few
being (with $\kappa^2 = \frac{1}{4}$)

\begin{eqnarray*}
u_2 & = & v_2 \,, \\
u_3 & = & v_3 + \frac{1}{2} v_{2,x} \,, \\
u_4 & = & v_4 + v_{3,x} + \frac{1}{4} v_{2,xx}\,.
\end{eqnarray*}

\noindent It is therefore conjectured that this
Moyal-KP hierarchy (with $\kappa^2=\frac{1}{4}$) is the same as the
KP hierarchy given by equation (2). Until this is proved, equation (8)
will be called the Moyal-KP hierarchy.
A similar approach was studied in {\bf [7]}, however the formalism used there
is slightly different and the $\kappa\rightarrow 0$ limit does not yield the
dKP equation directly, not without a scaling transformation, the
avoidance of which was one of the motivations of this section.

\section*{3. The reduction to the KdV hierarchy}

The KdV hierarchy may be obtained by imposing the constraint
${\cal L}\star{\cal L} = {\cal B}_2\,,$ and the evolution of $u_2$ is
given by

\[
{\partial {\cal B}_2\over \partial t_{2n+1} } =
\{ {\cal B}_{2n+1}, {\cal B}_2 \}_\kappa\,,
\]

\noindent all functions being independent of the even time variables. The
first couple of equations are given below, to show how the terms depend on
the parameter $\kappa\,:$

\[
\begin{array}{rcl}
u_{2,t_3} & = & \kappa^2 u_{2,xxx} + 3 u_2 u_{2,x} \,,         \\
u_{2,t_5} & = & 10 \kappa^4 u_{2,xxxxx} + 5\kappa^2 u_2 u_{2,xxx}
+ 10 \kappa^2 u_{2,x} u_{2,xx} + \frac{15}{2} u_2^2 u_{2,x} \,.
\end{array}
\]

\noindent In the limit as $\kappa\rightarrow 0$ one obtains the dispersionless
KdV hierarchy. Other $(1+1)$-dimensional hierarchies may be obtain by
imposing the appropriate constraints, as in the standard Sato theory.
Once again, a direct proof that this is
equivalent to the KdV hierarchy obtains using differential operators, as in
the Sato approach, is lacking.

\section*{4. The geometry of the Moyal-KP hierarchy}

A more geometrical way to describe the dKP hierarchy, equivalent to the
Lax equation (3), is to introduce a 2-form\footnote{In this section it will be
notationally convenient to set $x=t_1\,.$}

\[
\omega(\lambda) = \sum_{n=1}^\infty
d {\widetilde{\cal B}}_n\wedge dt_n\,,
\]

\noindent where ${\widetilde{\cal B}}_n = ({\cal L}^n)_{+}\,,$ i.e. the
$\kappa\rightarrow 0$
limit of the ${\cal B}_n\,.$
The dKP hierarchy then becomes the following conditions on the
2-form $\omega\,:$

\begin{equation}
\begin{array}{rcl}
\omega(\lambda)\wedge\omega(\lambda) & = & 0 \,, \\
d \omega(\lambda) & = & 0 \,.
\end{array}
\end{equation}

\noindent These equation imply, by Frobenious's theorem, the local existence
of functions ${\cal P}(\lambda)$ and ${\cal Q}(\lambda)$ such that
$\omega = d{\cal P}\wedge d{\cal Q}\,.$
In fact, one such pair of functions is given by

\begin{eqnarray*}
{\cal P}(\lambda) & = & {\cal L}\,,                                  \\
{\cal Q}(\lambda) & = & \sum_{n=1}^\infty n t_n {\cal L}^{n-1} \,,   \\
                  & \stackrel{\rm def}{=} & {\cal M}(\lambda)\,,
\end{eqnarray*}

\noindent and hence:

\bigskip

\noindent{\bf Proposition} {\bf [7]}

\bigskip

The dispersionless KP hierarchy is governed by the exterior differential
equation

\[
\omega = d{\cal L}\wedge d{\cal M}\,,
\]

\noindent with

\[ \{ {\cal L},{\cal M} \} =1\,. \]

To discuss the geometry of the KP hierarchy it is first convenient to redefine
the Moyal bracket

\[  \{ f,g\}_\kappa = f \star g - g \star f\,. \]

\noindent This amounts to rescaling the time variables, so now the limit
$\kappa\rightarrow 0$ is singular. The basic definitions of $\cal L$ and
${\cal B}_n$ remains unchanged.
In Sato theory the pseudo-differential operator

\[
W = 1 + \sum_{n=1}^\infty w_i \partial^{-n}
\]

\noindent plays a more fundamental role than the Lax operator $L$, and
the evolution of $W$ is governed by

\[
{\partial W\over\partial t_n} = B_n W - W \partial^n\,,
\]

\noindent where $L=W\partial W^{-1}\,$ and $B_n=(L^n)_{+}\,.$
{}From these equations it is straightforward to derive the Lax equation and
the zero curvature relations.
For the Moyal version of the KP hierarchy one may similarly define a function

\[ {\cal W} = 1 + \sum_{n=1}^\infty w_n \lambda^n \]

\noindent governed by

\[
{\partial {\cal W} \over\partial t_n} = {\cal B}_n \star {\cal W} -
{\cal W} \star \lambda^n\,.
\]

\noindent The Lax functions is then
${\cal L} = {\cal W}\star\lambda\star{\cal W}^{-1}$ (where ${\cal W}^{-1}$ is
defined uniquely by the relations
${\cal W}\star{\cal W}^{-1} = {\cal W}^{-1}\star{\cal W} = 1\,$), and
this satisfies the Lax equation (8).

\bigskip

Recently, another pseudo-differential operator was introduced in {\bf [8]},

\[ M = W (\sum_{n=1}^\infty n t_n \partial^{n-1}) W^{-1}\,. \]

\noindent this satisfying the equations

\begin{eqnarray*}
{\partial M\over \partial t_n} & = & [ B_n,M] \,. \\
{ [ L , M ] } & = & 1 \,.
\end{eqnarray*}

\noindent With this one may study the symmetries and other properties of the
KP hierarchy in terms
of the infinite dimensional Grassmannian manifold. Similarly, there is a
Moyal version of this operator:

\[
{\cal M}={\cal W}\star ( \sum_{n=1}^\infty n t_n \lambda^{n-1} )\star {\cal
W}^{-1}\,.
\]

\noindent This satisfies the relations

\begin{eqnarray*}
{\partial {\cal M}\over \partial t_n} & = & \{ {\cal B}_n,{\cal M} \}_\kappa
\,, \\
\{ {\cal L},{\cal M} \}_\kappa & = & 1 \,.
\end{eqnarray*}

\noindent This suggests that one may characterise solutions of the Moyal KP
hierarchy in terms of a Riemann-Hilbert problem in the Moyal loop group.

\bigskip

Another multidimensional integrable system that admits such a description is
the anti-self-dual Einstein equation {\bf [9]}.
These describe a complex 4-metric
with vanishing Ricci and anti-self-dual Weyl tensors. The metric may be
written in terms of a single function $\Omega\,,$ the K\"ahler potential, which
is governed by the equation

\[
\Omega_{,x{\tilde{x}}} \Omega_{,y{\tilde{y}}} -
\Omega_{,x{\tilde{y}}} \Omega_{,y{\tilde{x}}} = 1\,,
\]

\noindent or, using the Poisson bracket (4) (with respect to $\tilde{x}$ and
$\tilde{y}$ variables):

\begin{equation}
\{ \Omega_{,x},\Omega_{,y} \}  = 1\,.
\end{equation}

\noindent This equation (Plebanski's first heavenly equation {\bf [10]})
may also be
written in the form (10), with

\[
\omega(\lambda) = dx \wedge dy + \lambda
(\Omega_{x{\tilde{x}}} dx \wedge d{\tilde{x}} +
\Omega_{x{\tilde{y}}} dx \wedge d{\tilde{y}} +
\Omega_{y{\tilde{x}}} dy \wedge d{\tilde{x}} +
\Omega_{y{\tilde{y}}} dy \wedge d{\tilde{y}})+
\lambda^2 d{\tilde{x}}\wedge d{\tilde{y}}
\]

\noindent and the additional constraint $d\lambda=0\,.$ Once again one may
show the existence of function ${\cal P}(\lambda)$ and ${\cal Q}(\lambda)$
with $\omega = d{\cal P}\wedge d{\cal Q}$
and $\{ {\cal P}(\lambda),{\cal Q}(\lambda) \} = 1\,,$
connected by Riemann-Hilbert problems. A Moyal algebraic deformation of (11),
obtained by replacing the Poisson bracket by the Moyal bracket was
introduced in {\bf [11]}, and has been studied further by Takasaki {\bf [12]}
and Castro {\bf [13]}, the
former showing that it may be described in
terms of a Riemann-Hilbert problem in the corresponding Moyal loop group.

\section*{5. Comments}

The Moyal bracket was first introduced in an attempt to reformulate
quantum mechanics in terms of a distribution $f$ on phase space. From the
equation

\[ {\partial f\over \partial t} = \{ f, H\}_\kappa  \]

\noindent (together with an auxiliary equations for $f$) one can
derive a wavefunction satisfying the Schr\"odinger equation. Note that
the use of commutator relations is avoided. The theory outlined in this
paper is somewhat analogous; the use of differential operator and
commutator relations is replaced by the use of the Moyal bracket. Perhaps
these ideas may be useful in a proof of the conjecture that the hierarchy
(8) is the KP hierarchy.

\bigskip

It therefore seems likely that the entire theory of the KP hierarchy may be
reformulated in terms of the Moyal bracket and $\star$-products, thus totally
avoiding the use of pseudo-differential operators. One attraction of this
approach is that it is closer in spirit to the formulation of other
multidimensional integrable systems such as the anti-self-dual vacuum
equations. For this equation the Riemann-Hilbert problem may be used to define
an associated $3$-dimensional complex manifold known as twistor space. Such a
twistorial description of the KP equation has been sort for many years, after
the conjecture of Ward {\bf [14]} that all classical
integrable systems should admit
such a description. The failure to find this has lead to the
suggestion {\bf [15]} that a
more general version to twistor theory is needed to encompass systems such as
the KP equation. The use of the Moyal bracket in the study of the KP
hierarchy, as outlined in this paper, suggests that what might be required is
some sort of $\kappa$-deformed twistor space which would make use of the Moyal
bracket, rather than the Poisson bracket as used in conventional twistor
theory. This, however, remains pure speculation.

\section*{Acknowledgements}

Financial support was provided by the University of Newcastle, via the
Wilfred Hall fellowship.

\section*{References}

\begin{tabbing}

Space \= \kill

\\
\noindent{ \bf [1] } \> Y. Kodama, {\sl Phys. Lett. }{\bf A129} (1988) 223.\\
\> Y. Kodama and J. Gibbons, {\sl Phys. Lett. }{\bf A135} (1989) 167-170.\\

\\
\noindent{ \bf [2] } \> Y. Ohta, J. Satuma, D. Takahashi and T. Tokihiro,\\
\> {\sl Prog. Theor. Phys. Supp. }{\bf 94} (1988) 210. \\

\\
\noindent{ \bf [3] } \> J. Moyal, {\sl Proc. Cam. Phil. Soc. }{\bf 45} (1949)
99. \\

\\
\noindent{ \bf [4] } \> P. Fletcher, {\sl Phys. Lett. }{\bf B248} (1990) 323.\\
\> W. Arveson, {\sl Commun. Math. Phys. }{\bf 89} (1983) 77. \\

\\
\noindent{ \bf [5] } \> D.B. Fairlie, P. Fletcher and C.K. Zachos,
{\sl J. Math. Phys. }{\bf 31} (1990) 1085. \\

\\
\noindent{ \bf [6] } \> B.A. Kupershmidt, {\sl Lett. Math. Phys. }{\bf 20}
(1990) 19. \\

\\
\noindent{ \bf [7] } \> K. Takasaki and T. Takebe, {\sl Int. J. Mod. Phys.}
{\bf A7} {\sl Suppl. 1B} (1992), 889.\\

\\
\noindent{ \bf [8] } \> P.G Grinevich and A.Yu. Orlov, in {\sl Problems of
Modern Quantum Field Theory} \\
\> eds. A.A. Belavin, A.U. Klimyk and A.B. Zamalodchikov (Springer-Verlag,
1989). \\

\\
\noindent{ \bf [9] } \> R. Penrose, {\sl Gen. Rel. Grav. }{\sl 7} (1976) 31.\\

\\
\noindent{ \bf [10] } \> J.F. Plebanski, {\sl J. Math. Phys. }{\bf 16} (1975)
2395.\\

\\
\noindent{ \bf [11] } \> I.A.B. Strachan, {\sl Phys. Lett. }{\bf B283} (1992)
63.\\

\\
\noindent{ \bf [12] } \> K. Takasaki, {\sl Dressing operator approach to Moyal
algebraic deformation of} \\
\> {\phantom{K. Takasaki, }}{\sl self dual gravity}, Kyoto preprint,
KUCP-0054/92,\\
\> {\phantom{K. Takasaki, }}{\sl Nonabelian KP hierarchy with Moyal
algebraic coefficients},\\
\> {\phantom{K. Takasaki, }}to appear {\sl J. Geom. Phys. }{\bf 14}. \\

\\
\noindent{ \bf [13] } \> C. Castro, {\sl Nonlinear $W_\infty$ algebras from
nonlinear integrable deformations of} \\
\> {\phantom{C. Castro, }}{\sl self dual gravity}, preprint I.A.E.C.-4-94.\\
\> {\phantom{C. Castro, }}{\sl A Universal $W_\infty$ algebra and
quantization of integrable deformations} \\
\> {\phantom{C. Castro, }}{\sl of self-dual gravity}, preprint I.A.E.C.-2-94.\\

\\
\noindent{ \bf [14] } \> R.S. Ward, {\sl Phil. Trans. R. Soc. }{\bf A315}
(1985) 451,\\
\> {\phantom{R.S. Ward, }}in {\sl Twistors in Mathematics and Physics},
eds. T.N. Bailey and \\
\> {\phantom{R.S.Ward, }} R.J. Baston (Cambridge University Press, 1990). \\

\\
\noindent{ \bf [15] } \> L.J. Mason, {\sl Generalized twistor correspondences,
d-bar problems and the}\\
\> {\phantom{L.J. Mason, }}{\sl KP equations},to appear in the proceedings of
the
Twistor Theory \\
\> {\phantom{L.J. Mason, }}conference held in Newton Abbot, 1993. \\

\end{tabbing}

\end{document}